\pdfoutput=1

\RequirePackage{fix-cm}

\documentclass[twocolumn]{svjour3}

\smartqed

\usepackage{graphicx,amssymb}
\usepackage{epsf,psfig}
\usepackage{txfonts}
\usepackage{natbib}
\usepackage[unicode]{hyperref}

\hypersetup{pdftitle = The title of my PDF, pdfauthor = My name, pdfsubject= The subject, pdfkeywords = keyword1 keyword2 keyword3}
\hypersetup{colorlinks = true, linkcolor = blue, anchorcolor = red, citecolor = blue, filecolor = red, pagecolor = red, urlcolor = blue}

\journalname{Astrophysics and Space Science}

\begin{document}

\title{Unveiling the influence of the radiation pressure in nature of orbits in the photogravitational restricted three-body problem}

\author{Euaggelos E. Zotos}

\institute{Department of Physics, School of Science, \\
Aristotle University of Thessaloniki, \\
GR-541 24, Thessaloniki, Greece \\
Corresponding author's email: {evzotos@physics.auth.gr}}

\date{Received: 27 July 2015 / Accepted: 16 September 2015 / Published online: 1 October 2015}

\titlerunning{Influence of the radiation pressure in nature of orbits in the photogravitational restricted three-body problem}

\authorrunning{Euaggelos E. Zotos}

\maketitle

\begin{abstract}
The case of the planar circular photogravitational restricted three-body problem where the more massive primary is an emitter of radiation is numerically investigated. A thorough numerical analysis takes place in the configuration $(x,y)$ and the $(x,C)$ space in which we classify initial conditions of orbits into three main categories: (i) bounded, (ii) escaping and (iii) collisional. Our results reveal that the radiation pressure factor has a huge impact on the character of orbits. Interpreting the collisional motion as leaking in the phase space we related our results to both chaotic scattering and the theory of leaking Hamiltonian systems. We successfully located the escape as well as the collisional basins and we managed to correlate them with the corresponding escape and collision times. We hope our contribution to be useful for a further understanding of the escape and collision properties of motion in this interesting version of the restricted three-body problem.


\end{abstract}

\section{Introduction}
\label{intro}

Stars (like our Sun) exert not only gravitation, but also radiation pressure on bodies moving nearby. At the same time, it is well known that dust particles are characterized by a considerable sailing capacity (cross-section to mass ratio), and hence are subject to a sizable effect of light pressure from the star, being one of the possible mechanisms for the formation and evolution of gas-dust clouds. Therefore the study of motion of particles under photogravitation effect becomes realistic an important field of research.

Unfortunately the classical planar circular restricted three-body problem is not valid for investigating the motion of material points in the solar system where the third mass has considerable sailing capacity (for example cosmic dust, stellar wind, etc). Thus it is reasonable to modify the classical model by superposing a radiative repulsion field, whose source coincides with the source of the gravitational field (the Sun), onto the gravitational field of the main body. Several modifications of the restricted three-body problem have been proposed for the study of the motion of a massless particle in the Solar System (see, e.g., \citet{SMB85,SSR76}). These modifications include additional forces in the potential of the classical problem which may make it more realistic for certain applications. \citet{P03} has stated that particles, such as small meteors or cosmic dust, are comparably affected by gravitational and light radiation forces as they approach luminous celestial bodies. The importance of the radiation influence on celestial bodies has been recognized by many scientists, especially in connection with the formation of concentrations of interplanetary and interstellar dust or grains in planetary and binary star systems, as well as the perturbations on artificial satellites.

The term photo-gravitational restricted three-body problem was introduced by \citet{R50}. This extended version of the classical restricted three-body problem takes into account only the radiation pressure component of the radiation drag, which is the next most powerful component after the gravitational forces. Later, \citet{R53} performed a complete treatment of the behavior of the equilibrium points. In both papers, however, Radzievskii, who was primarily interested in the solar problem, only treated a limited range of radiation pressure parameters (in particular when only one massive body is luminous) and did not consider the question of the linear stability of the equilibrium points. Moreover Radzievskii investigated that besides five libration points of classical problem, there exist out-of-plane equilibrium points $L_6$ and $L_7$. \citet{C70} extended his work by introducing the Pointing-Robertson effect. Many authors (i.e., \citet{P76,S80,SMB85,M94,R01,DNMY09}) developed and extended this work to introduce more understandable issues related to the motion of particle in the field of radiating primaries.

In the previous years several authors have made studies on Lagrangian points in the restricted three-body problem by considering the more massive primary or both primaries as source of radiation. Some of the important contributions are by \citet{KP78,BC79,KT85,L88,T94,RZ88a,RZ88b,XLY94,KPR06,P06}. \citet{S87} and \citet{SI95} investigated the linear stability of the triangular points in the planar case by considering the more massive primary as source of radiation and the smaller primary as oblate spheroid and found that the critical mass value decreases with the increase in radiation pressure. \citet{K96} studied libration solutions of the photogravitational restricted three-body problem. Recently \citet{ARS06,ARS08} studied the combined effects of perturbations due to coriolis forces, centrifugal forces, radiation pressure and oblateness on the linear stability of the triangular libration points.

Another interesting topic in the photogravitational restricted three-body problem is the computation of families of periodic orbits (i.e., \citet{KPR06,KPP08,Per03,PKM02,PKD08,PPK12}. In the same vein, \citet{MAB09} have studied periodic orbits generated by Lagrangian solutions of the restricted three-body problem when the bigger body is a source of radiation and the
smaller is an oblate spheroid. At this point we should emphasize, that all the above-mentioned references on periodic orbits in the photogravitational restricted three-body problem are exemplary rather than exhaustive, taking into account that a vast quantity of related literature exists.

In a recent paper \citet{Z15b} we explored how the oblateness coefficient influences the orbital content in the restricted three-body problem. In this work we shall use this paper as a guide following the same numerical techniques in order to unveil how the radiation pressure factor affects the character of orbits in the photogravitational restricted three-body problem. The structure of the paper is as follows: In Section \ref{mod} we outline the properties of the considered mathematical model along with some necessary theoretical details. All the computational methods we used in order to determine the nature of the orbits are described in Section \ref{cometh}. In the following Section, we conduct a thorough numerical investigation revealing the overall orbital structure (bounded regions and basins of escape/collision) of the system and how it is affected by the value of the radiation pressure with respect to the total orbital energy. Our paper ends with Section \ref{conc}, where the discussion and the conclusions of this work are given.

\section{Presentation of the mathematical model}
\label{mod}

It would be very illuminating to recall the basic properties and some aspects of the planar circular photogravitational restricted three-body problem. The two primaries move on circular orbits with the same Kepler frequency around their common center of gravity, which is assumed to be fixed at the origin of the coordinates. The third body (test particle) moves in the same plane under the gravitational field of the two primaries. The mass of the infinitesimal test body is supposed to be so small that it's influence on the motion of the primaries is negligible. We assume that the bigger primary is an intense emitter of radiation (replicating the scenario of the Solar system) which contributes for the effect of radiation pressure, while the second smaller primary is not radiating. It should be pointed out that, in contrast to the classical three-body problem, in the photogravitational problem the force acting on the particle depends not only on the parameters of the radiating body (temperature, size, density, etc) but also on the parameters of the particle itself (size, density, etc).

The photogravitational version of the restricted three-body problem is derived in a similar way to the classical problem \citep{S67}. In fact, we adopt the notation and formulation used in \citet{SMB85}. The units of length, mass and time are taken so that the sum of the masses, the distance between the primaries and the angular velocity is unity, which sets the gravitational constant $G = 1$. A rotating rectangular system whose origin is the center of mass of the primaries and whose $Ox$-axis contains the primaries is used. The mass parameter is $\mu = m_2/(m_1 + m_2)$, where $m_1 = 1 - \mu$ and $m_2 = \mu$ are the masses of the primaries with $m_1 > m_2$, such that $m_1 + m_2 = 1$. In our study we consider an intermediate value of the mass parameter, that is $\mu = 1/5$. This value remains constant throughout the paper.

We have fixed the center of mass at (0,0), while the centers $C_1$ and $C_2$ of the two primary bodies are located at $(-\mu, 0)$ and $(1-\mu,0)$, respectively. The forces experienced by the test particle in the coordinate system rotating with angular velocity $\omega = 1$ and origin at the center of the mass are according to \citep{S67,S82} derivable from the following total time-independent effective potential function
\begin{equation}
\Omega(x,y) =  \frac{q (1 - \mu)}{r_1} + \frac{\mu}{r_2} + \frac{1}{2}\left(x^2  + y^2 \right),
\label{pot}
\end{equation}
where $(x,y)$ are the coordinates of the test particle, while
\[
r_1 = \sqrt{\left(x + \mu\right)^2 + y^2},
\]
\begin{equation}
r_2 = \sqrt{\left(x + \mu - 1\right)^2 + y^2},
\end{equation}
are the position vectors from the larger and smaller primaries to the test particle, respectively. We decided not to include the oblateness coefficient thus considering both primaries as spherically symmetric bodies in order to focus our study on only one variable parameter $(q)$ which controls the intensity of the radiation.

The parameter $q$ is the radiation pressure factor which is defined as $q = 1 - \beta$, where $\beta = F_p/F_g$ is the ratio of the solar radiation pressure force $(F_p)$ to the gravitational attraction force $(F_g)$ of the bigger primary \citep{SMB85}. It should be noted that $q$ is a reduction factor which depends on the size and the shape of the third body (see e.g., \citet{S80}). For a general body, such as a planet, $q \sim 1$. For a small body on the other hand, such as an asteroid or a satellite, $1 < q < 1$, while for extremely small objects (i.e., a dust grain) it is possible that $q < 0$. In general terms the value of $q$ could be taken in the interval $(- \infty, 1]$ \citep{ZY93}. In this work we restrict our investigation in the interval $q \in (0, 1]$.

The scaled equations of motion describing the motion of the test body in the dimensionless corotating frame read
\[
\ddot{x} - 2\dot{y} = \frac{\partial \Omega(x,y)}{\partial x},
\]
\begin{equation}
\ddot{y} + 2\dot{x} = \frac{\partial \Omega(x,y)}{\partial y}.
\label{eqmot}
\end{equation}
The dynamical system (\ref{eqmot}) admits the well know Jacobi integral
\begin{equation}
J(x,y,\dot{x},\dot{y}) = 2\Omega(x,y) - \left(\dot{x}^2 + \dot{y}^2 \right) = C,
\label{ham}
\end{equation}
where $\dot{x}$ and $\dot{y}$ are the velocities, while $C$ is the Jacobi constant which is conserved and defines a three-dimensional invariant manifold in the total four-dimensional phase space. Thus, an orbit with a given value of it's energy integral is restricted in its motion to regions in which $C \leq 2\Omega(x,y)$, while all other regions are forbidden to the test body. The value of the energy $E$ is related with the Jacobi constant by $C = - 2E$.

The dynamical system has five equilibria known as Lagrangian points \citep{S67} at which
\begin{equation}
\frac{\partial \Omega(x,y)}{\partial x} = \frac{\partial \Omega(x,y)}{\partial y} = 0.
\label{lps}
\end{equation}
Three of them, $L_1$, $L_2$, and $L_3$, are collinear points located in the $x$-axis. The central stationary point $L_1$ is a local minimum of the potential function $\Omega(x,y)$. The stationary points $L_2$ and $L_3$ are saddle points. Let $L_2$ located at $x > 0$, while $L_3$ be at $x < 0$. The points $L_4$ and $L_5$ on the other hand, are local maxima of the gravitational potential, enclosed by the banana-shaped isolines. The Lagrangian points are very important especially for astronautical applications. This can be seen in the Sun-Jupiter system where several thousand asteroids (collectively referred to as Trojan asteroids), are in orbits of equilibrium points. In practice these Lagrangian points have proven to be very useful indeed since a spacecraft can be made to execute a small orbit about one of these equilibrium points with a very small expenditure of energy (see e.g., \citet{SL14}).

\begin{figure*}[!ht]
\centering
\resizebox{\hsize}{!}{\includegraphics{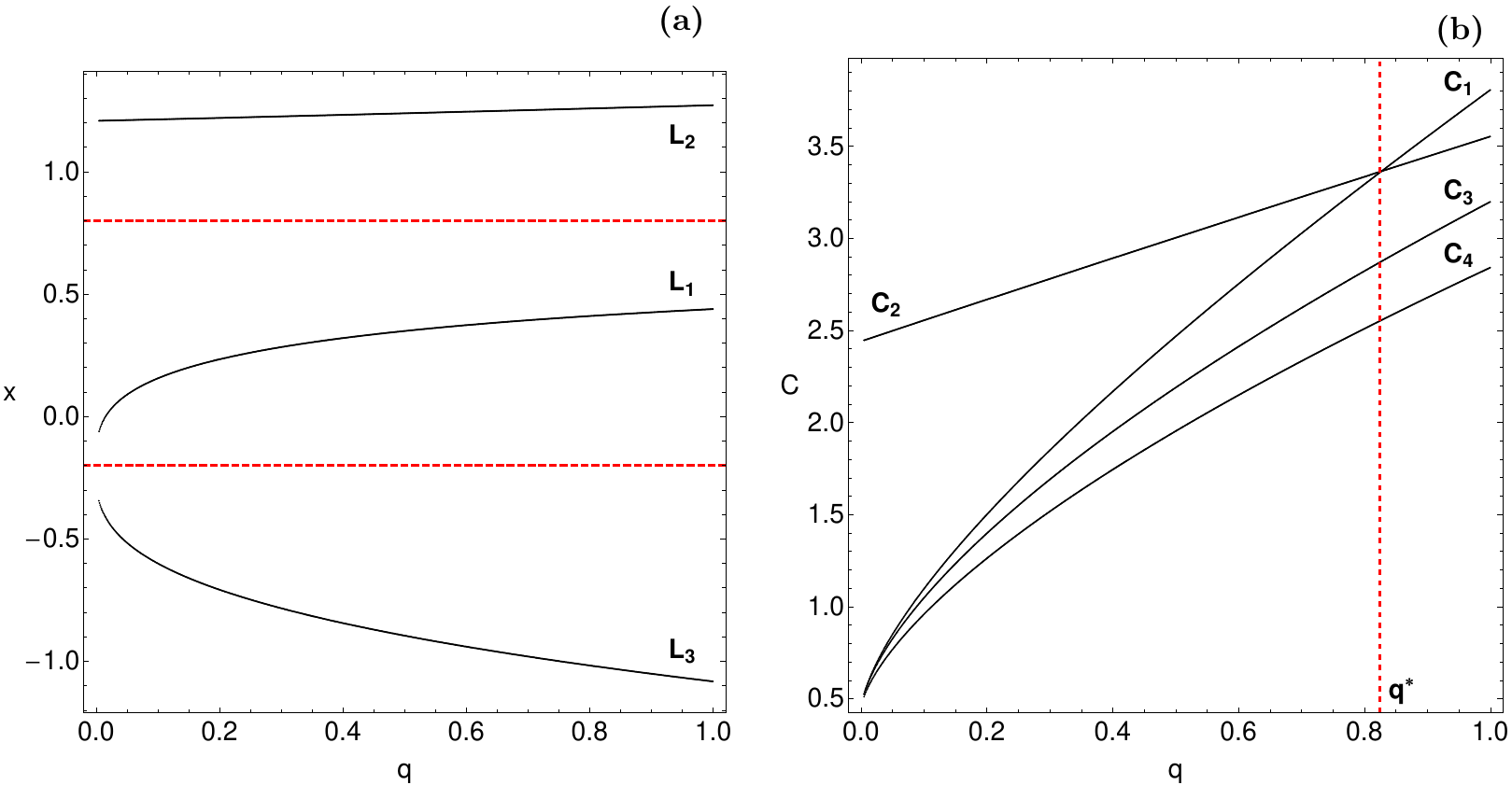}}
\caption{(a-left): Evolution of the position of the Lagrangian points as a function of the radiation pressure factor $q$. The horizontal red dashed lines indicate the position of the centers of the two primaries. (b-right): Evolution of the critical Jacobi constant values as a function of the radiation pressure factor $q$. The vertical red dashed line indicate the $q^{*}$ value where $C_1 = C_2$.}
\label{theor}
\end{figure*}

\begin{figure*}[!tH]
\centering
\resizebox{\hsize}{!}{\includegraphics{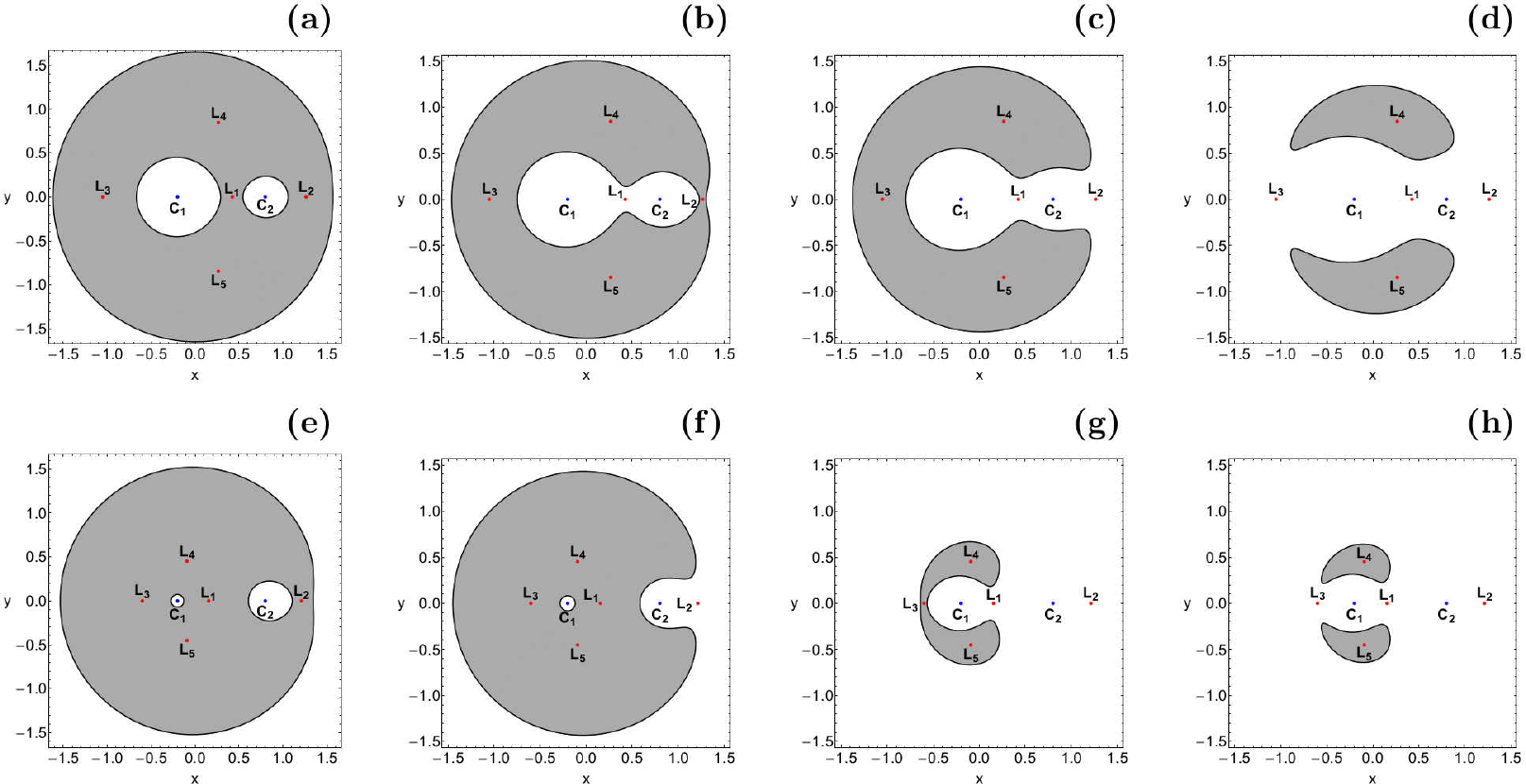}}
\caption{The first four Hill's regions configurations for the photogravitational RTBP system when (a-d): $q = 0.9$ and (e-h): $q = 0.1$. The white domains correspond to the Hill's regions, gray shaded domains indicate the forbidden regions, while the thick black lines depict the Zero Velocity Curves (ZVCs). The red dots pinpoint the position of the Lagrangian points, while the positions of the centers of the two primary bodies are indicated by blue dots. (a): $C = 3.80$; (b): $C = 3.45$; (c): $C = 3.30$; (d): $C = 2.95$; (e): 2.65; (f): 2.40; (g): 1.05; (h): 1.03.}
\label{conts}
\end{figure*}

The position of the Lagrangian points is actually a function of the radiation pressure factor $q$. In Fig. \ref{theor}a we show how the position of the collinear points $L_1$, $L_2$ and $L_3$ evolves with respect to the value of the radiation pressure factor $q$. It is seen that the collinear points $L_1$, $L_2$ and $L_3$ come nearer to the radiating primary with the increase in the radiation pressure (remember that the radiation pressure increases when the value of $q$ decreases). In particular it is found that the collinear points $L_1$ and $L_3$ coincide with the center of the radiating primary when $q = 0$. In the same vein, \citet{NRS08} studied the influence of the oblateness and the radiation pressure on angular frequencies at collinear points. The Jacobi constant values at the Lagrangian points $L_i, i = 1, ..., 5$ are denoted by $C_i$ and are critical values (note that $C_4 = C_5$). Fig. \ref{theor}b shows the evolution of the $C_i$ critical values as a function of the radiation pressure factor. We see that $C_1$, $C_3$ and $C_4$ evolve very similarly, while $C_2$ exhibits a linear evolution. It is evident that $C_1$, $C_3$ and $C_4$ tend to a common value when $q \rightarrow 0$, or in other words, when the more massive primary body is an intense emitter of radiation. Furthermore, when $q = q^{*} = 0.82450170$ the values of $C_1$ and $C_2$ coincide, while $C_2 > C_1$ for $0 < q < q^{*}$.

The projection of the four-dimensional phase space onto the configuration (or position) space $(x,y)$ is called the Hill's regions and is divided into three domains: (i) the interior region for $x(L_3) \leq x \leq x(L_2)$; (ii) the exterior region for $x < x(L_3)$ and $x > x(L_2)$; (iii) the forbidden regions. The boundaries of these Hill's regions are called Zero Velocity Curves (ZVCs) because they are the locus in the configuration $(x,y)$ space where the kinetic energy vanishes. The structure of the Hill's regions strongly depends on the value of the Jacobi constant and also on the value of the radiation pressure factor. There are five distinct cases regarding the Hill's regions:
\begin{itemize}
  \item $C > C_1$: All necks are closed, so there are only bounded and collisional basins. When $q < q^{*}$ and as the heavier primary becomes more and more radiating the allowed region around this primary is considerably reduced.
  \item $C_2 < C < C_1$: When $q^{*} < q <1$ only the neck around $L_1$ is open thus allowing orbits to move around both primaries. For $q < q^{*}$ the neck around $L_1$ is closed, while the neck around $L_2$ is open thus allowing orbits to escape from the interior region.
  \item $C_3 < C < E_2$: The neck around $L_2$ is open, so orbits can enter the exterior region and escape form the system. For $q < q^{*}$ the area of the forbidden regions is much smaller than in the case when $q^{*} < q <1$. In particular the forbidden regions when $q < q^{*}$ are surrounding the domain only around the radiating primary.
  \item $C_4 < C < C_3$: The necks around both $L_2$ and $L_3$ are open, therefore orbits are free to escape through two different escape channels.
  \item $C < C_4$: The banana-shaped forbidden regions disappear, so motion over the entire configuration $(x,y)$ space is possible.
\end{itemize}
In Fig. \ref{conts}(a-h) we present the structure of the first four possible Hill's region configurations for $q = 0.9$ (a-d) and $q = 0.1$ (e-h). We observe in Figs. \ref{conts}d and \ref{conts}h the two openings (exit channels) at the Lagrangian points $L_2$ and $L_3$ through which the test body can enter the exterior region and then leak out. In fact, we may say that these two exits act as hoses connecting the interior region of the system where $x(L_3) \leq x \leq x(L_2)$ with the ``outside world" of the exterior region. In Table \ref{table1} we provide the location of the Lagrangian points and the critical values of the Jacobi constant when $q = \{0.1, 0.5, 0.9\}$.

\begin{table*}[!ht]
\begin{center}
   \caption{The position of the Lagrangian points and the critical values of the Jacobi constant for three values of the radiation pressure factor $q$.}
   \label{table1}
   \setlength{\tabcolsep}{10pt}
   \begin{tabular}{@{}lcccc}
      \hline
      $q$ & $L_1$ & $L_2$ & $L_3$ & $L_4$ \\
      \hline
      0.9 & (0.42472220, 0) & (1.26416203, 0) & (-1.05146290, 0) & (0.26608487, 0.84553807) \\
      0.5 & (0.34849544, 0) & (1.23766062, 0) & (-0.89578805, 0) & (0.11498026, 0.72852450) \\
      0.1 & (0.15484417, 0) & (1.21294280, 0) & (-0.60056059, 0) & (-0.09227826, 0.45148587) \\
      \hline
      $q$ & $C_1$ & $C_2$ & $C_3$ & $C_4$ \\
      \hline
      0.9 & 3.551290549 & 3.443371379 & 3.012826587 & 2.677207404 \\
      0.5 & 2.465911264 & 3.002213547 & 2.188090188 & 1.951905259 \\
      0.1 & 1.094884116 & 2.553126233 & 1.045713142 & 0.957064325 \\
      \hline
   \end{tabular}
\end{center}
\end{table*}

\section{Computational methods and criteria}
\label{cometh}

The motion of the test third body is restricted to a three-dimensional surface $C = const$, due to the existence of the Jacobi integral. With polar coordinates $(r,\phi)$ in the center of the mass system of the corotating frame the condition $\dot{r} = 0$ defines a two-dimensional surface of section, with two disjoint parts $\dot{\phi} < 0$ and $\dot{\phi} > 0$. Each of these two parts has a unique projection onto the configuration $(x,y)$ space. In order to explore the orbital structure of the system we need to define samples of initial conditions of orbits whose properties will be identified. For this purpose, we define for several values of the Jacobi constant $C$, as well as for the radiation pressure factor $q$ dense uniform grids of $1024 \times 1024$ initial conditions regularly distributed on the configuration $(x,y)$ space inside the area allowed by the value of the Jacobi constant. Following a typical approach, the orbits are launched with initial conditions inside a certain region, called scattering region, which in our case is a square grid with $-2\leq x,y \leq 2$.

\begin{figure}[!tH]
\centering
\includegraphics[width=\hsize]{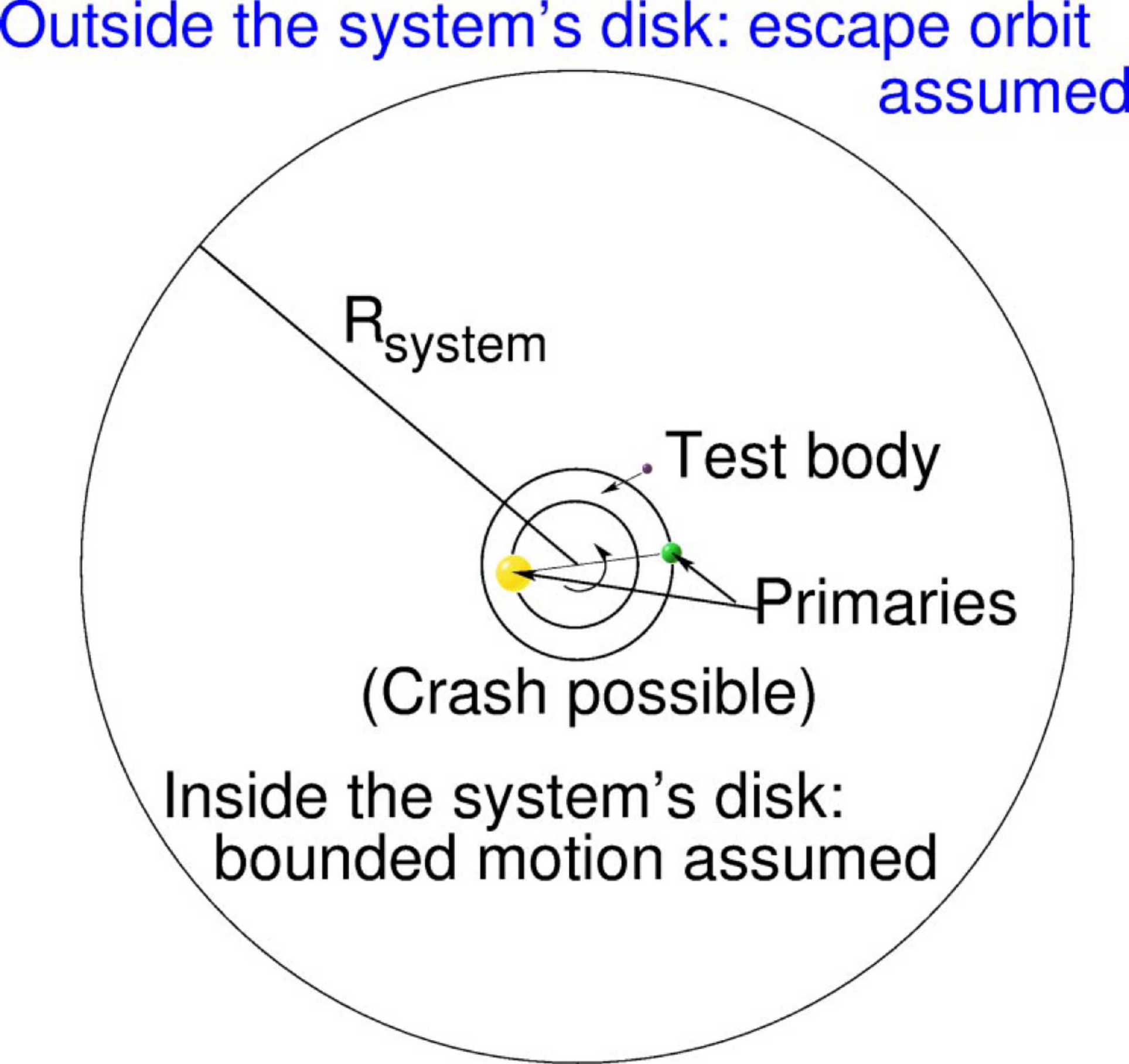}
\caption{Schematic picture of the three different types of motion. The motion is considered to be bounded if the test body stays confined for integration time $t_{\rm max}$ inside the system's disk with radius $R_d = 10$, while the motion is unbounded and the numerical integration stops when the test body crosses the system's disk with velocity pointing outwards. Collision with one of the primaries occurs when the test body crosses the disk of radii $R_{m1}$ and $R_{m2}$ of the primaries.}
\label{crit}
\end{figure}

In the photogravitational RTBP system the configuration space extends to infinity thus making the identification of the type of motion of the test body for specific initial conditions a rather demanding task. There are three possible types of motion for the test body: (i) bounded motion around one of the primaries, or even around both; (ii) escape to infinity; (iii) collision into one of the two primaries. Now we need to define appropriate numerical criteria for distinguishing between these three types of motion. The motion is considered as bounded if the test body stays confined for integration time $t_{\rm max}$ inside the system's disk with radius $R_d$ and center coinciding with the center of mass origin at $(0,0)$. Obviously, the higher the values of $t_{\rm max}$ and $R_d$ the more plausible becomes the definition of bounded motion and in the limit $t_{\rm max} \rightarrow \infty$ the definition is the precise description of bounded motion in a finite disk of radius $R_d$. Consequently, the higher these two values, the longer the numerical integration of initial conditions of orbits lasts. In our calculations we choose $t_{\rm max} = 10^4$ and $R_d = 10$ (see Fig. \ref{crit}) as in \citet{N04,N05} and \citet{Z15a,Z15b}. We decided to include a relatively high disk radius $(R_d = 10)$ in order to be sure that the orbits will certainly escape from the system and not return back to the interior region. Furthermore, it should be emphasized that for low values of $t_{\rm max}$ the fractal boundaries of stability islands corresponding to bounded motion become more smooth. Moreover, an orbit is identified as escaping and the numerical integration stops if the test body intersects the system's disk with velocity pointing outwards at a time $t_{\rm esc} < t_{\rm max}$. Finally, a collision with one of the primaries occurs if the test body, assuming it is a point mass, crosses the disk with radius $R_m$ around the primary. For the larger oblate primary we choose $R_{m_1} = 10^{-4}$. Generally, it is assumed that the radius of a celestial body (e.g., a planet) is directly proportional to the cubic root of its mass. Therefore, for the sake of simplicity of the numerical calculations we adopt the simple relation between the radii of the primaries
\begin{equation}
R_{m_2} = R_{m_1} \times \left(2\mu\right)^{1/3},
\label{radii}
\end{equation}
which was introduced in \citet{N05}. In \citet{N04,N05} it was shown that the radii of the primaries influence the area of collision and escape basins.

As it was stated earlier, in our computations, we set $10^4$ time units as a maximum time of numerical integration. The vast majority of escaping orbits (regular and chaotic) however, need considerable less time to escape from the system (obviously, the numerical integration is effectively ended when an orbit moves outside the system's disk and escapes). Nevertheless, we decided to use such a vast integration time just to be sure that all orbits have enough time in order to escape. Remember, that there are the so called ``sticky orbits" which behave as regular ones during long periods of time. Here we should clarify, that orbits which do not escape after a numerical integration of $10^4$ time units are considered as non-escaping or trapped.

The equations of motion (\ref{eqmot}) for the initial conditions of all orbits are forwarded integrated using a double precision Bulirsch-Stoer \verb!FORTRAN 77! algorithm (e.g., \citet{PTVF92}) with a small time step of order of $10^{-2}$, which is sufficient enough for the desired accuracy of our computations. Here we should emphasize, that our previous numerical experience suggests that the Bulirsch-Stoer integrator is both faster and more accurate than a double precision Runge-Kutta-Fehlberg algorithm of order 7 with Cash-Karp coefficients. Throughout all our computations, the Jacobian energy integral (Eq. (\ref{ham})) was conserved better than one part in $10^{-11}$, although for most orbits it was better than one part in $10^{-12}$. For collisional orbits where the test body moves inside a region of radius $10^{-2}$ around one of the primaries the Lemaitre's global regularization method is applied.

\section{Numerical results \& Orbit classification}
\label{numres}

The main numerical task is to classify initial conditions of orbits in the $\dot{\phi} < 0$ part\footnote{We choose the $\dot{\phi} < 0$ instead of the $\dot{\phi} > 0$ part simply because in \citet{Z15a} we seen that it contains more interesting orbital content.} of the surface of section $\dot{r} = 0$ into three categories: (i) bounded orbits; (ii) escaping orbits and (iii) collisional orbits. Moreover, two additional properties of the orbits will be examined: (i) the time-scale of collision and (ii) the time-scale of the escapes (we shall also use the terms escape period or escape rates). In this work we shall explore these dynamical quantities for various values of the total orbital energy, as well as for the radiation pressure factor $q$. In particular, we choose four energy levels which correspond to the last four Hill's regions configurations. The first Hill's regions configuration contain only bounded and collisional orbits around the two primaries, so the orbital content is not so interesting.

At this point we would like to clarify how the initial conditions of orbits are generated in the $\dot{\phi} < 0$ part of the surface of section $\dot{r} = 0$. The conditions $\dot{\phi} < 0$ and $\dot{r} = 0$ in polar coordinates along with the existence of the Jacobi integral of motion (\ref{ham}) suggest that the four initial conditions of orbits in cartesian coordinates are
\begin{eqnarray}
x &=& x_0, \ \ \ y = y_0, \nonumber\\
\dot{x_0} &=& \frac{y_0}{r}\sqrt{2\Omega(x,y) - C}, \nonumber\\
\dot{y_0} &=& - \frac{x_0}{r}\sqrt{2\Omega(x,y) - C},
\end{eqnarray}
where $r = \sqrt{x_0^2 + y_0^2}$. In the following color-coded grids (or orbit type diagrams - OTDs) each pixel is assigned a color according to the orbit type. Thus the initial conditions of orbits on the $(x,y)$-plane are classified into bounded orbits, unbounded or escaping orbits and collisional orbits. In this special type of Poincar\'{e} Surface of Section (PSS) the phase space emerges as a close and compact mix of escape basins, collisional basins and stability regions. Our numerical calculations indicate that apart from the escaping and collisional orbits there is also a considerable amount of non-escaping orbits. In general terms, the majority of non-escaping regions corresponds to initial conditions of regular orbits, where an adelphic integral of motion is present, restricting their accessible phase space and therefore hinders their escape.

In the following we are going to explore the orbital content of the configuration $(x,y)$ space in three different cases regarding the intensity of the radiation of the more massive primary body. In every case we choose such values of the Jacobi constant $C$ so that the shape of the Hill's region configuration to be similar, as much as possible, in all three cases under consideration.

\subsection{Case I: Low radiating primary}

\begin{figure*}[!tH]
\centering
\resizebox{\hsize}{!}{\includegraphics{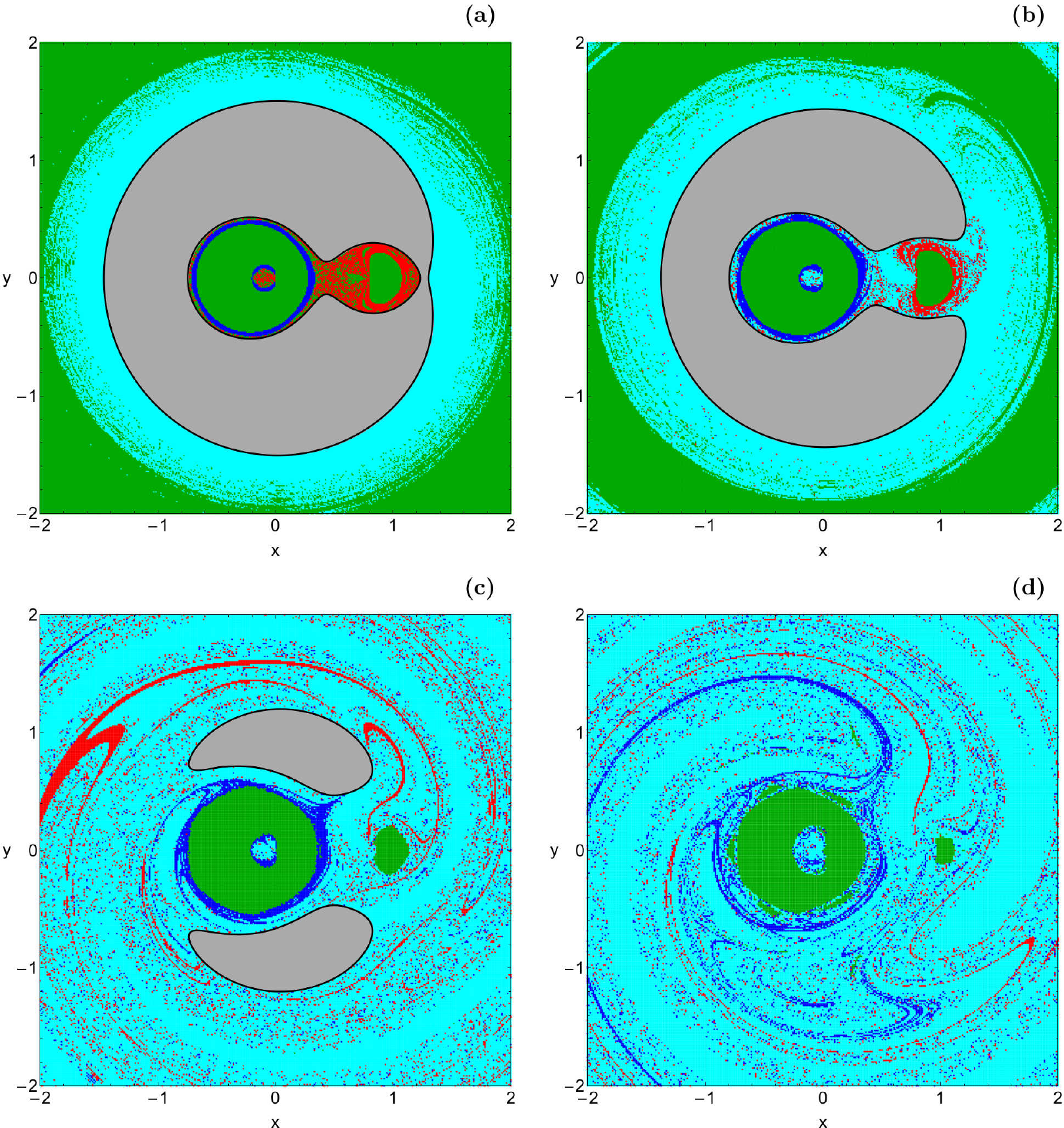}}
\caption{The orbital structure of the $\dot{\phi} < 0$ part of the surface of section $\dot{r} = 0$ when $q = 0.9$. (a-upper left): $C = 3.45$; (b-upper right): $C = 3.30$; (c-lower left): $C = 2.90$; (d-lower right): $C = 2.50$. The color code is the following: bounded orbits (green), collisional orbits to radiating primary 1 (blue), collisional orbits to primary 2 (red), and  escaping orbits (cyan).}
\label{Rl}
\end{figure*}

\begin{figure*}[!tH]
\centering
\resizebox{\hsize}{!}{\includegraphics{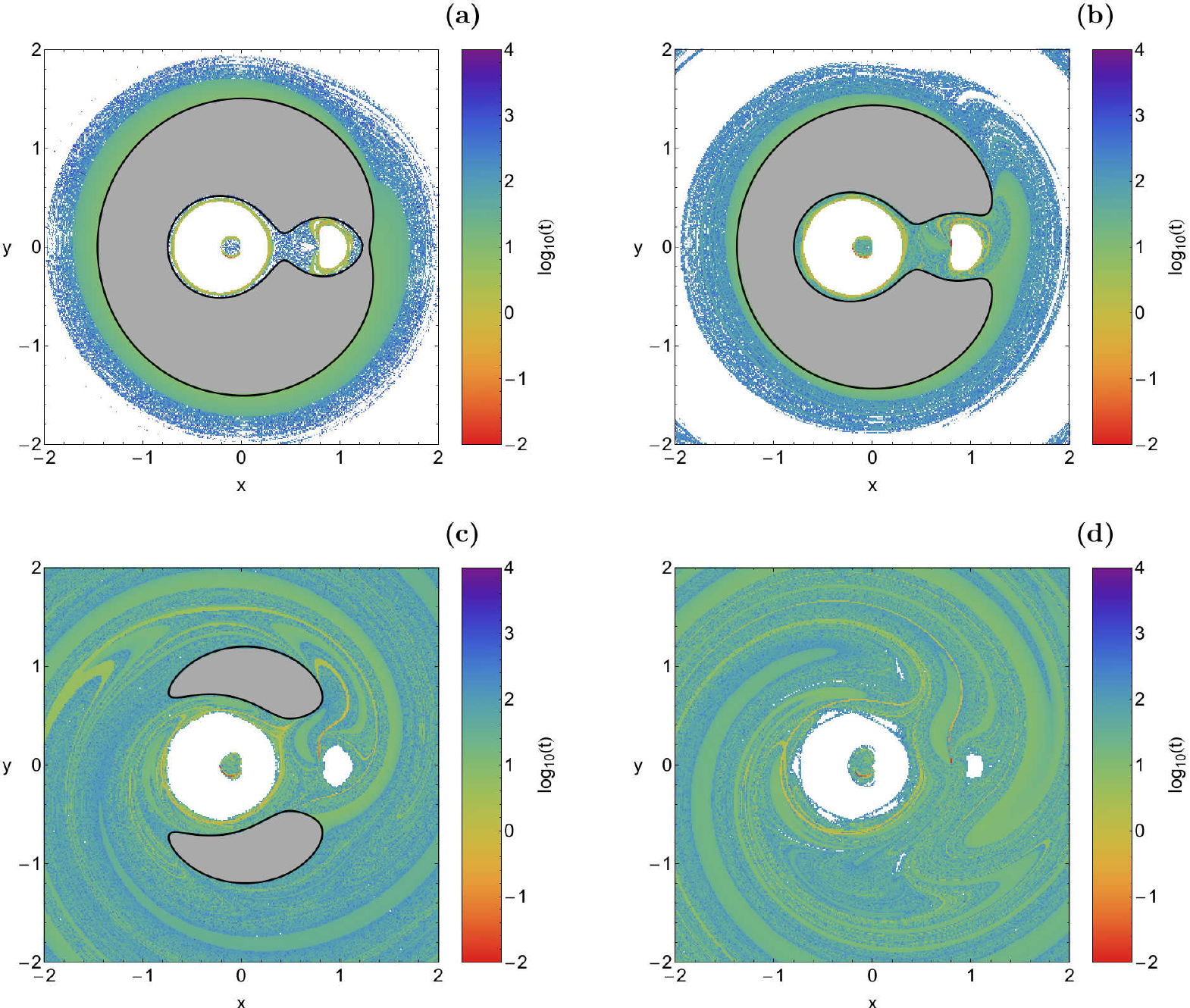}}
\caption{Distribution of the escape and collisional time of the orbits on the $\dot{\phi} < 0$ part of the surface of section $\dot{r} = 0$ when $q = 0.9$ for the values of the Jacobi constant of Fig. \ref{Rl}. The darker the color, the larger the escape/collisional time. Initial conditions of bounded regular orbits are shown in white.}
\label{Rlt}
\end{figure*}

Our first case under investigation concerns a low radiating primary where the radiation pressure factor is $q = 0.9$. In Fig. \ref{Rl}(a-d) the OTD decompositions of the $\dot{\phi} < 0$ part of the surface of section $\dot{r} = 0$ reveal the orbital structure of the configuration $(x,y)$ space for four values of the Jacobi constant $C$. The black solid lines denote the ZVC, while the inaccessible forbidden regions are marked in gray. The color of a point represents the orbit type of a test body which has been launched with pericenter position at $(x,y)$. When $C = 3.45$ which corresponds to the second type of Hill's region configurations we observe in Fig. \ref{Rl}a the followings: (i) around the centers of the two primaries there are stability islands, while in the exterior region another ring-shaped stability island is present. The stability islands located in the interior region correspond to regular orbits around one of the primary bodies, while the stability island in the exterior region is formed by initial conditions of regular orbits that circulate around both primaries; (ii) The region in the boundaries of the stability islands of the interior region is mainly occupied by orbits which lead to collision. Moreover, the collision basin around primary 2 (non radiating) is stronger than that around primary 1; (iii) Near the center of primary 1 we can identify a small hole which contains a mixture of collisional and bounded orbits; (iv) outside the forbidden region there is a well-formed circular basin of escaping orbits. When $C = 3.30$ the neck around $L_2$ opens and therefore initial conditions of escaping orbits emerge in the interior region. We see in Fig. \ref{Rl}b that the orbital structure around the more massive radiating primary does not practically change apart from the fact that escaping orbits are added in the mixture of orbits in the small hole around center $C_1$. Around the smaller non radiating primary on the other hand, it is seen that the collisional basin is smaller, while initial conditions of collisional orbits to primary 2 start to leak out to the exterior region. In Fig. \ref{Rl}c where $C = 2.90$ both necks around $L_2$ and $L_3$ are open now. Once more the orbital content around the radiating primary remains almost unperturbed. A minor change is the increase of the central hole. The collisional basin around primary 2 has completely disappear, while in the exterior region we can identify a strong presence of initial conditions of orbits which collide to primary 2 and form well defined spiral basins. On the other hand, the initial conditions of orbits in the exterior region which collide to radiating primary 1 do not form collisional basins but they appear as delocalized scattered points. In addition, the stability island contains initial conditions of orbits that circulate around both primaries is absent now. The banana-shaped forbidden regions disappear in Fig. \ref{Rl}d where $C = 2.50$. The stability islands around the centers of the two primaries are still present however, the island around primary 2 is smaller with respect to the previous three energy cases. Furthermore, the ring-shaped collisional basin around the stability island of primary 1 is now destabilized and the corresponding initial conditions form thin spiral bands in the configuration $(x,y)$ space. In addition, the collisional basins to primary 2 are now weaker with respect to that shown earlier in Fig. \ref{Rl}c. Additional numerical calculation (not shown here) suggest that for lower values of the Jacobi constant, or in other words for higher values of the total orbital energy, the fractality of the configuration space increases.

In the following Fig. \ref{Rlt} we show how the escape and collisional times of orbits are distributed on the configuration $(x,y)$ space for the four values of the Jacobi constant discussed in Fig. \ref{Rl}(a-d). Light reddish colors correspond to fast escaping/collional orbits, dark blue/purple colors indicate large escape/collional times, while white color denote stability islands of regular motion. Note that the scale on the color bar is logarithmic. Inspecting the spatial distribution of various different ranges of escape time, we are able to associate medium escape time with the stable manifold of a non-attracting chaotic invariant set, which is spread out throughout this region of the chaotic sea, while the largest escape time values on the other hand, are linked with sticky motion around the stability islands of the two primary bodies. As for the collisional time we see that a small portion of orbits with initial conditions very close to the vicinity of the centers of the primaries collide with them almost immediately, within the first time steps of the numerical integration. Looking more carefully Fig. \ref{Rlt}d we clearly observe that when $C = 2.5$ the area of the stability region around primary 2 is indeed smaller with respect to the tree previous cases shown in Figs. \ref{Rlt}(a-c). Moreover, around the stability island of primary 1 we identify a chain of five small stability islands that correspond to secondary resonant orbits. Thus we may say that high enough values of the total orbital energy influence also the stability regions around both primaries.

\subsection{Case II: Intermediate radiating primary}

\begin{figure*}[!tH]
\centering
\resizebox{\hsize}{!}{\includegraphics{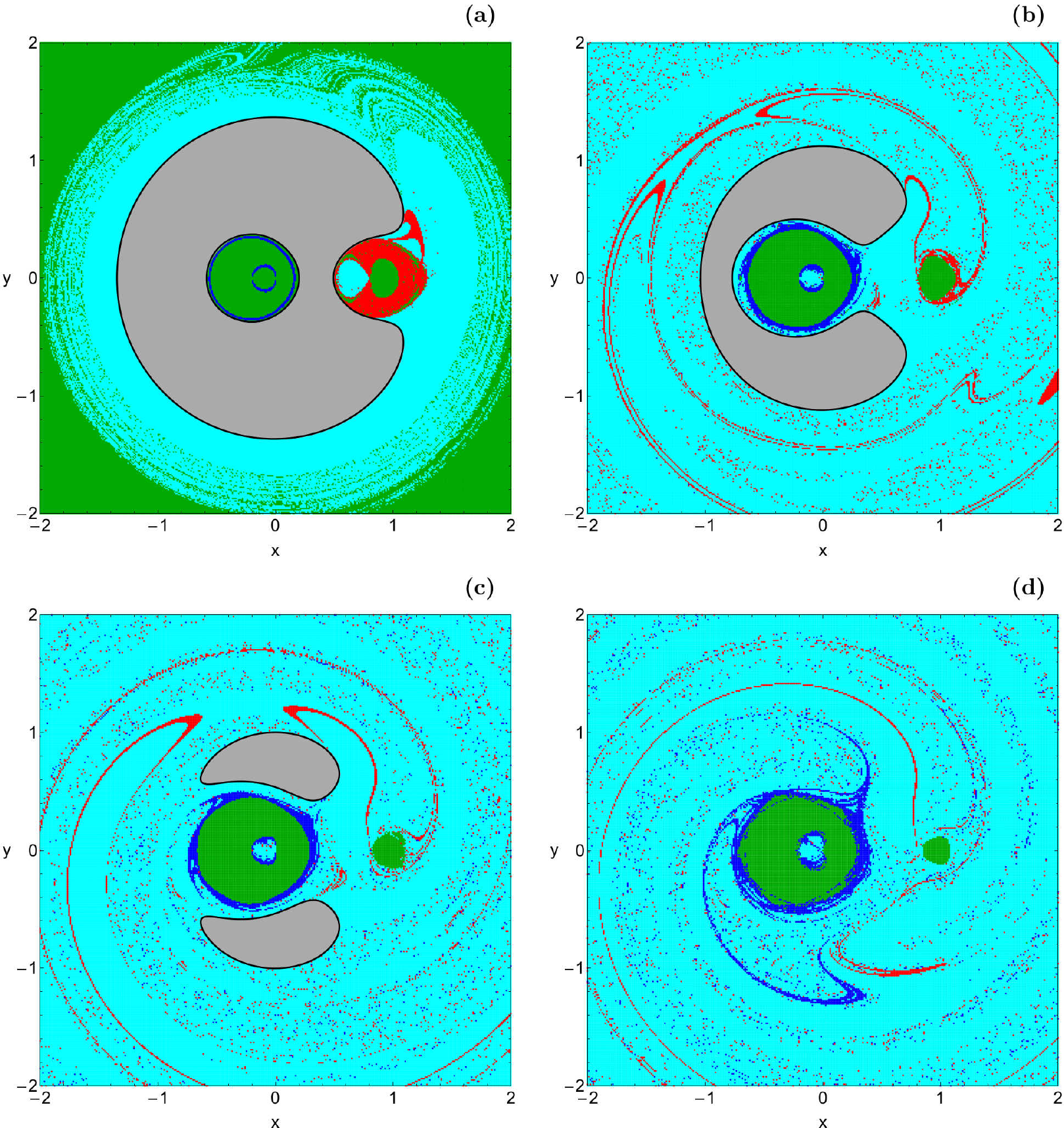}}
\caption{The orbital structure of the $\dot{\phi} < 0$ part of the surface of section $\dot{r} = 0$ when $q = 0.5$. (a-upper left): $C = 2.70$; (b-upper right): $C = 2.25$; (c-lower left): $C = 2.10$; (d-lower right): $C = 1.90$. The color code is the same as in Fig. \ref{Rl}.}
\label{Rm}
\end{figure*}

\begin{figure*}[!tH]
\centering
\resizebox{\hsize}{!}{\includegraphics{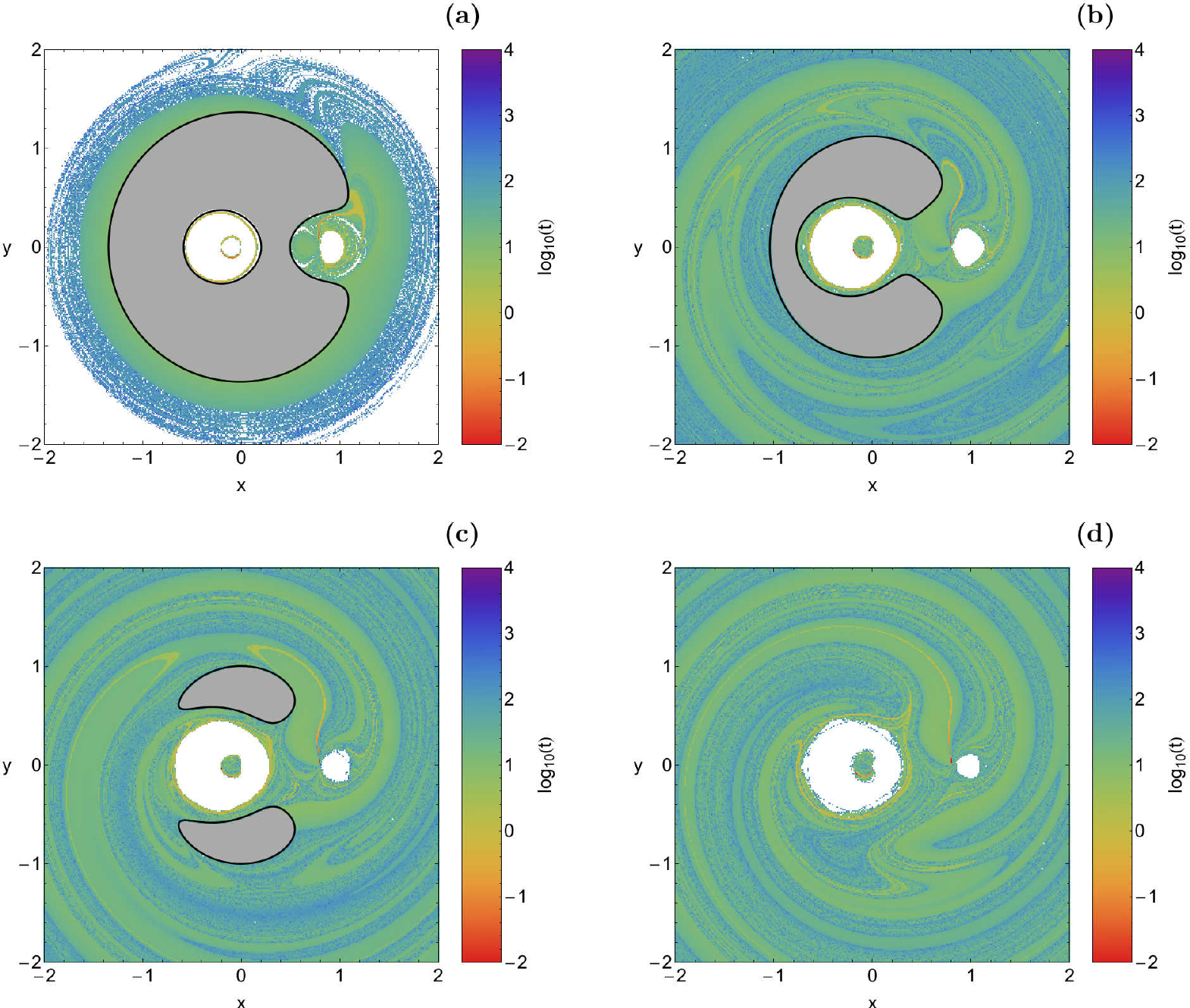}}
\caption{Distribution of the escape and collisional time of the orbits on the $\dot{\phi} < 0$ part of the surface of section $\dot{r} = 0$ when $q = 0.5$ for the values of the Jacobi constant of Fig. \ref{Rm}.}
\label{Rmt}
\end{figure*}

We continue our exploration considering the case where we have an intermediate radiating primary with radiation pressure factor $q = 0.5$. The orbital structure of the configuration $(x,y)$ space is unveiled in Fig. \ref{Rm} through the OTD decompositions of the $\dot{\phi} < 0$ part of the surface of section $\dot{r} = 0$. When $C = 2.70$ it is seen in Fig. \ref{Rm}a that the throat around $L_1$ is closed. The orbital structure around the more massive radiating primary is almost the same as the one discussed earlier in Fig. \ref{Rl}a. We have to point out however, that in this case the hole near the center $C_1$ is somehow altered since we observe a thin ring-shaped collisional basin inside the regular domain. Once more, around primary 2 there is a well-defined collisional basin, while at the left side of the center $C_2$ there is a small escaping basin. As for the exterior region there are no significant differences with respect to the $q = 0.9$ case shown in Fig. \ref{Rl}a. In Fig. \ref{Rm}b we present the orbital structure of the configuration space when $C = 2.25$. In the exterior region we see several collisional basins to primary 2 forming complicated shapes inside the vast escaping domain. Our computations strongly suggest that initial conditions of collisional orbits to radiating primary 1 are very rare in the exterior region. Things are quite similar in Fig. \ref{Rm}c where $C = 2.10$. Now the collisional basin around the stability island of primary 2 is absent, while at the same time the collisional basins to primary 2 in the exterior region are weaker with respect to the previous case of Fig. \ref{Rm}b. Moreover, the amount of initial conditions of orbits that collide to primary 1 located in the exterior region is greater. Looking at Fig. \ref{Rm}d, where $C = 1.90$, it becomes evident that as we move to lower values of the Jacobi constant, or in other words to greater values of the total orbital energy, the presence in the exterior region of collisional basins to primary 2 weakens, while the presence of collisional basins to radiating primary 1 is strengthened. Indeed at low enough values of $C$ the exterior region is a highly fractal mixture of escaping and collisional initial conditions of orbits.

The distribution of the escape and collisional times of orbits on the configuration space is shown in Fig. \ref{Rmt}. One may observe that the results are very similar to those presented earlier in Fig. \ref{Rlt}, where we found that orbits with initial conditions inside the escape and collisional basins have the smallest escape/collision rates, while on the other hand, the longest escape/collisional rates correspond to orbits with initial conditions in the fractal regions of the OTDs. Our calculations reveal, and this can be seen better in Figs. \ref{Rmt}(a-d), that also in this case the value of the Jacobi constant affects the size of the stability island of motion around non-radiating primary 2.

\subsection{Case III: Intense radiating primary}

\begin{figure*}[!tH]
\centering
\resizebox{\hsize}{!}{\includegraphics{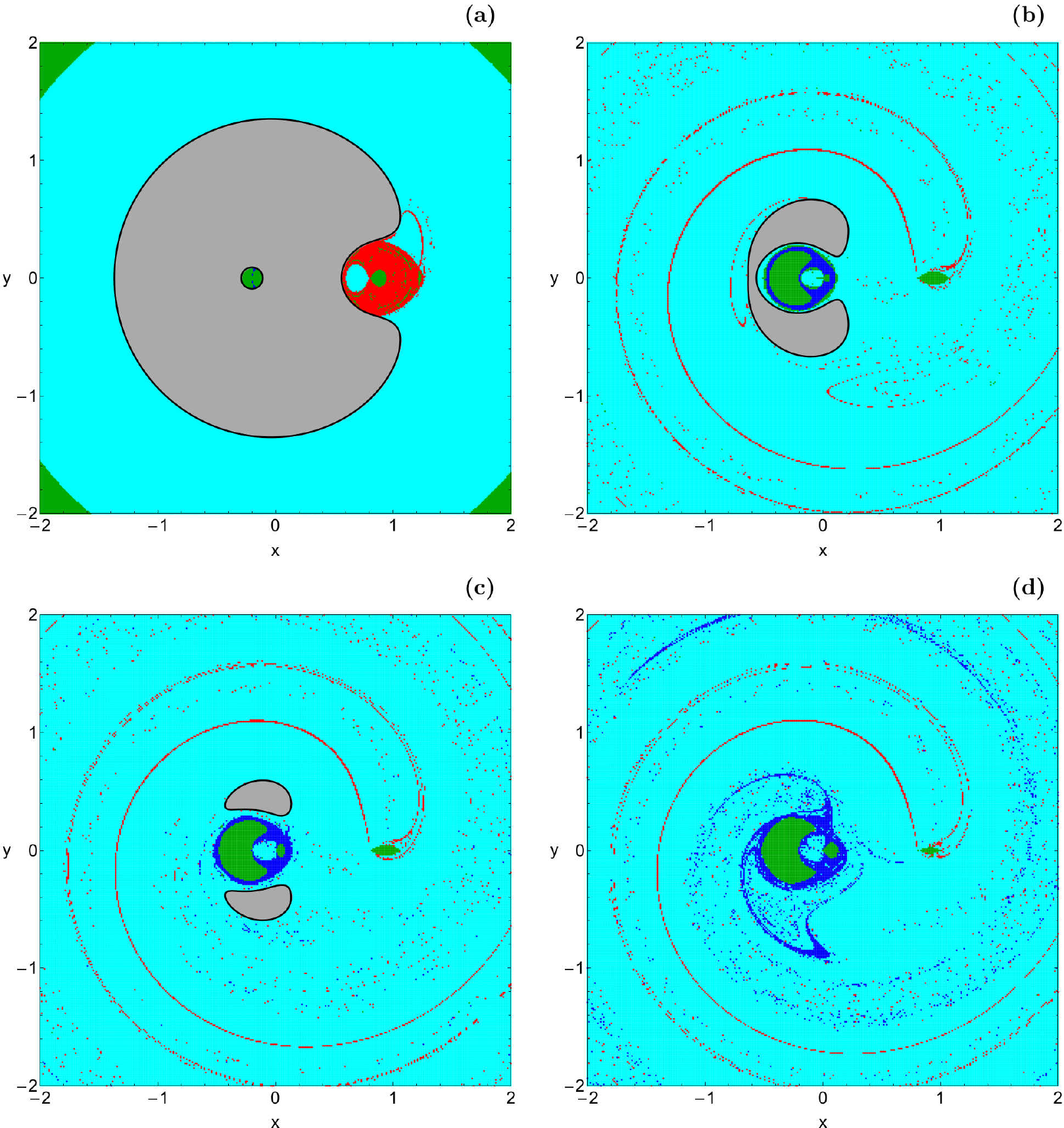}}
\caption{The orbital structure of the $\dot{\phi} < 0$ part of the surface of section $\dot{r} = 0$ when $q = 0.1$. (a-upper left): $C = 2.20$; (b-upper right): $C = 1.05$; (c-lower left): $C = 1.00$; (d-lower right): $C = 0.90$. The color code is the same as in Fig. \ref{Rl}.}
\label{Rh}
\end{figure*}

\begin{figure*}[!tH]
\centering
\resizebox{\hsize}{!}{\includegraphics{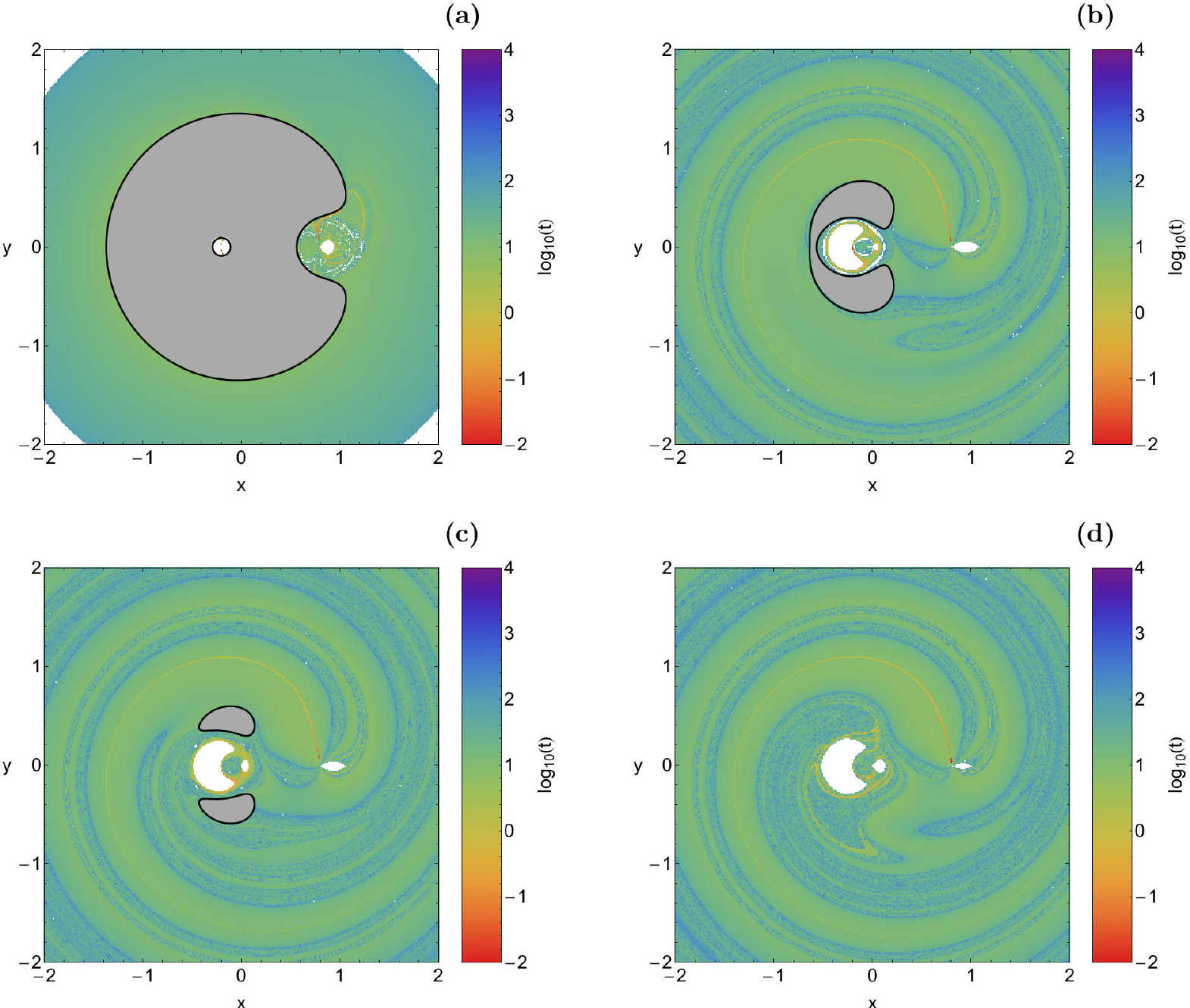}}
\caption{Distribution of the escape and collisional time of the orbits on the $\dot{\phi} < 0$ part of the surface of section $\dot{r} = 0$ when $q = 0.1$ for the values of the Jacobi constant of Fig. \ref{Rh}.}
\label{Rht}
\end{figure*}

The last case under consideration involves the scenario when the more massive primary is an intense radiating body with radiation pressure factor $q = 0.1$. Once more, all the different aspects of the numerical approach remain exactly the same as in the two previously studied cases. Fig. \ref{Rh} presents the orbital structure of the configuration space through the OTD decompositions of the $\dot{\phi} < 0$ part of the surface of section $\dot{r} = 0$. In Fig. \ref{Rh}a where it corresponds to the second type of Hill's region configuration for $C = 2.20$ we observe that the allowed area around primary 1 is extremely small, while the orbital structure around primary 2 is in general terms similar to that shown earlier in Fig. \ref{Rm}a. In the exterior region there is a solid escape basin and after that the stability region of regular orbits that circulate around both primaries. In contrast to the previous two cases the boundaries between the two domains of the exterior region are extremely smooth. When $C = 1.05$ it is seen in Fig. \ref{Rh}b that the forbidden regions embrace the area around only primary 1. Inside the forbidden regions there is a stability island, the usual hole and a ring-shaped collisional basin. In the exterior region we identified initial conditions of orbits that collide to primary 2, while collisional orbits to primary 1 do not have initial conditions in the exterior region. When the forbidden regions break into two symmetrical parts for $C = 1.00$ we observe in Fig. \ref{Rh}c that the orbital structure around primary 1 remains almost the same, while the thin spiral collisional basins to primary 2 are still present in the exterior region. However now several initial conditions of orbits that collide to primary 1 are present as scattered lonely points in the exterior region. Finally in Fig. \ref{Rh}d where $C = 0.90$ one can see that at the right side of the center of radiating primary 1 there is also a second smaller stability island. Furthermore, now the collisional orbits to primary 1 form collisional basins in the exterior region, while the collisional basins to primary 2 seem to weaken.

In Fig. \ref{Rht} we depict the distribution of the escape and collisional times of orbits on the configuration space. One can see similar outcomes with that presented in the two previous subsections. At this point, we would like to emphasize that the basins of escape can be easily distinguished in Fig. \ref{Rht}, being the regions with intermediate colors indicating fast escaping orbits. Indeed, our numerical computations suggest that orbits with initial conditions inside these basins need no more than 10 time units in order to escape from the system. Furthermore, the collisional basins are shown with reddish colors where the corresponding collisional time is less than one time unit of numerical integration. We observe that the second stability island at the right side of the primary 1 it emerges rather quickly from the third type of Hill's region configurations. It should be also noted that the area of the stability island around primary 2 decreases with decreasing $C$ or increasing value of the total orbital energy.

The OTDs shown in Figs. \ref{Rl}, \ref{Rm} and \ref{Rh} have both fractal and non-fractal (smooth) boundary regions which separate the escape basins from the collisional basins. Such fractal basin boundaries is a common phenomenon in leaking Hamiltonian systems (e.g., \citet{BGOB88,dML99,dMG02,STN02,ST03,TSPT04}). In the photogravitational RTBP system the leakages are defined by both escape and collision conditions thus resulting in three exit modes. However, due to the high complexity of the basin boundaries, it is very difficult, or even impossible, to predict in these regions whether the test body (e.g., a satellite, asteroid, planet etc) collides with one of the primary bodies or escapes from the dynamical system.

\subsection{An overview analysis}

\begin{figure*}[!tH]
\centering
\resizebox{\hsize}{!}{\includegraphics{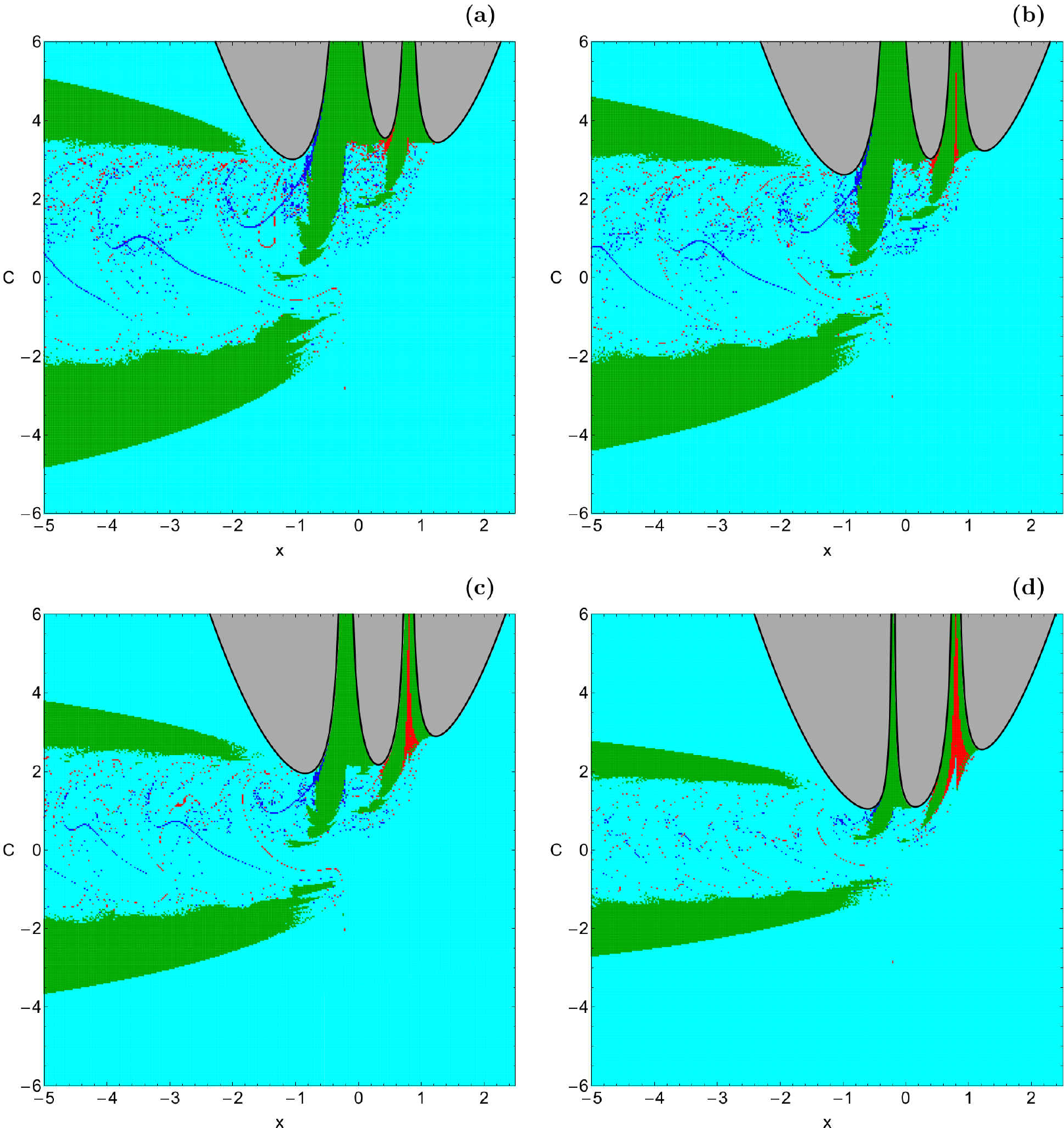}}
\caption{Orbital structure of the $(x,C)$ plane when (a-upper left): $q = 0.90$; (b-upper right): $q = 0.7$; (c-lower left): $q = 0.4$; (d-lower right): $q = 0.1$. The color code is the same as in Fig. \ref{Rl}.}
\label{xC}
\end{figure*}

\begin{figure*}[!tH]
\centering
\resizebox{\hsize}{!}{\includegraphics{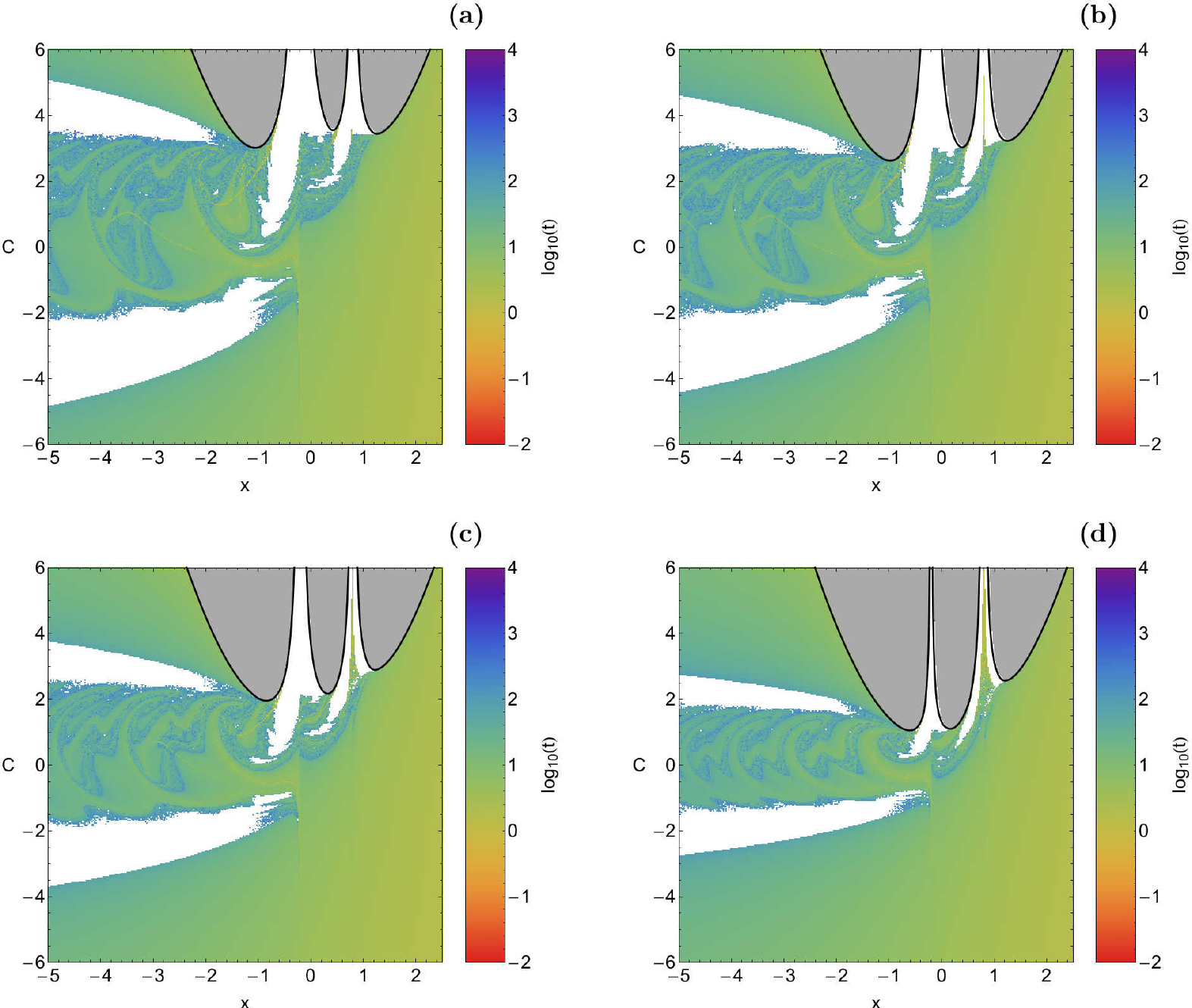}}
\caption{Distribution of the escape and collisional time of the orbits on the $(x,C)$ plane for the values of the radiation pressure factor of Fig. \ref{xC}.}
\label{xCt}
\end{figure*}

\begin{figure*}[!tH]
\centering
\resizebox{\hsize}{!}{\includegraphics{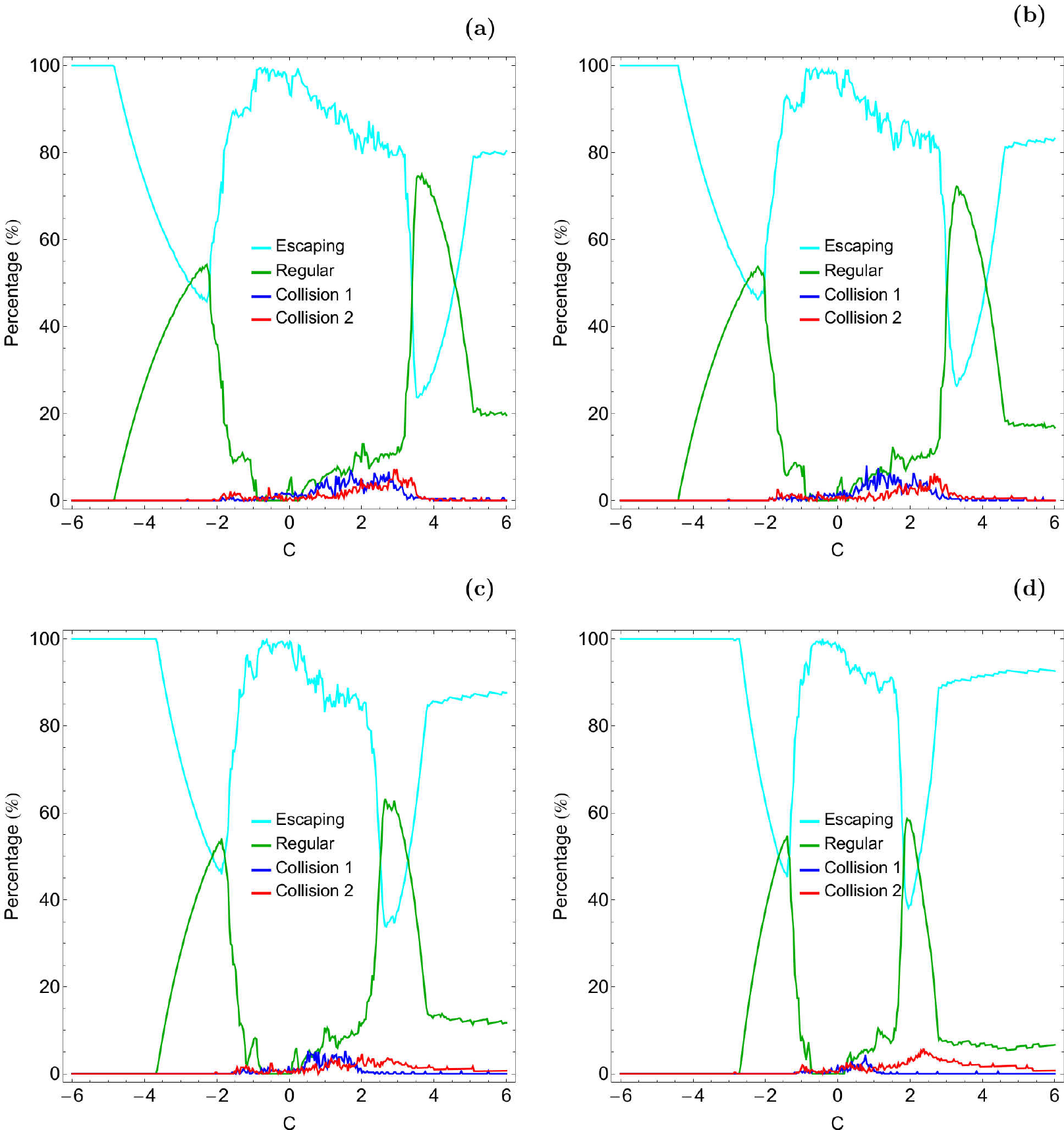}}
\caption{Evolution of the percentages of escaping, regular and collisional orbits on the $(x,C)$-plane as a function of the value of the Jacobi constant $C$. (a-upper left): $q = 0.9$; (b-upper right): $q = 0.7$; (c-lower left): $q = 0.4$; (d-lower right): $q = 0.1$.}
\label{pxC}
\end{figure*}

\begin{figure}[!tH]
\includegraphics[width=\hsize]{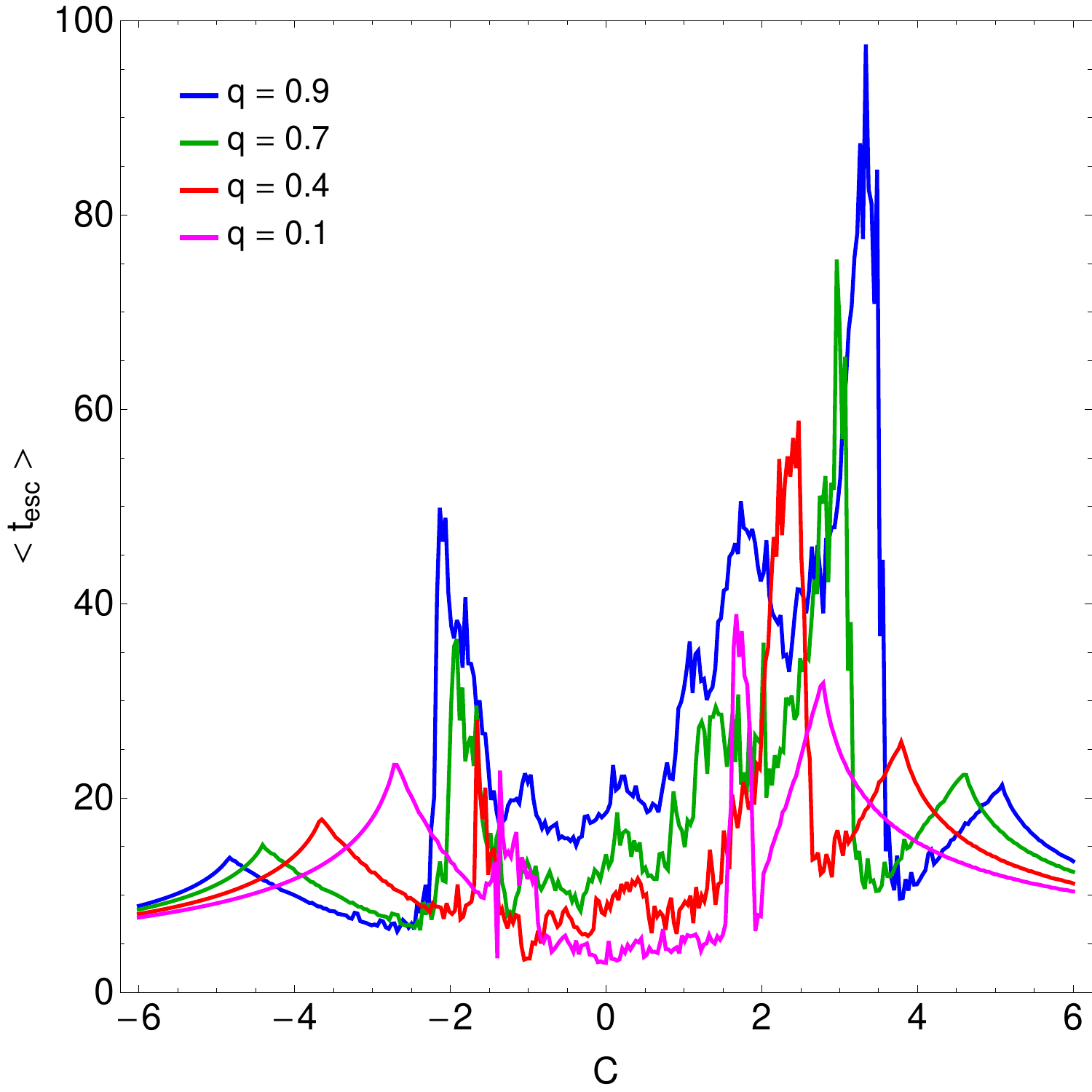}
\caption{Evolution of the average escape time of orbits $< t_{\rm esc} >$ as a function of the value of the Jacobi constant $C$.}
\label{tesc}
\end{figure}

\begin{figure*}[!tH]
\centering
\resizebox{\hsize}{!}{\includegraphics{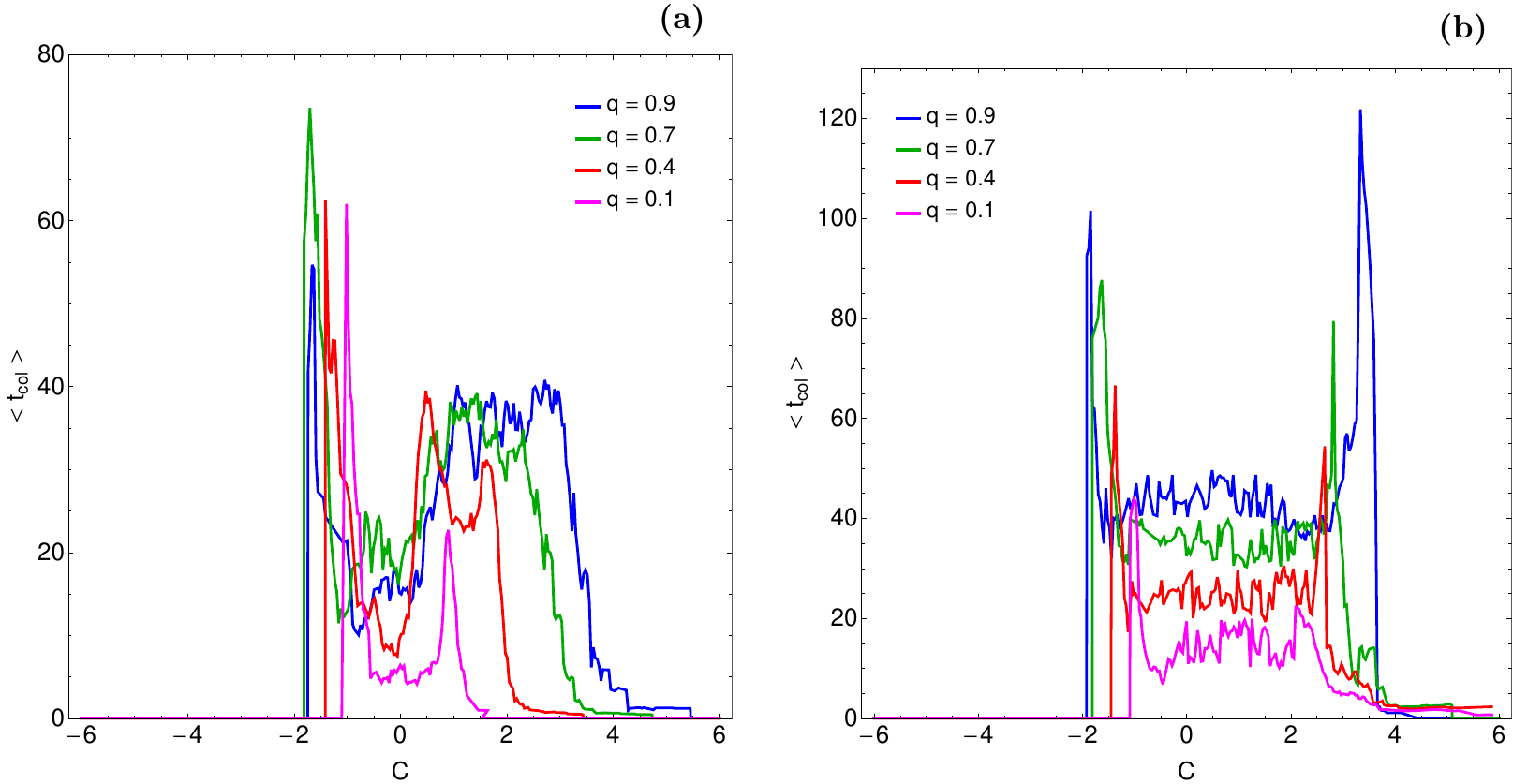}}
\caption{(a-left): The average collision time of orbits $< t_{\rm col} >$ (a-left): to radiating primary 1 and (b-right): to primary 2 as a function of the value of the Jacobi constant $C$.}
\label{tcol}
\end{figure*}

The color-coded OTDs in the configuration $(x,y)$ space provide sufficient information on the phase space mixing however, for only a fixed value of the Jacobi constant (or the total orbital energy) and also for orbits that traverse the surface of section retrogradely. H\'{e}non \citep{H69}, introduced a new type of plane which can provide information not only about stability and chaotic regions but also about areas of bounded and unbounded motion using the section $y = \dot{x} = 0$, $\dot{y} > 0$ (see also \citet{BBS08}). In other words, all the initial conditions of the orbits of the test particles are launched from the $x$-axis with $x = x_0$, parallel to the $y$-axis $(y = 0)$. Consequently, in contrast to the previously discussed types of planes, only orbits with pericenters on the $x$-axis are included and therefore, the value of the Jacobi constant $C$ can now be used as an ordinate. In this way, we can monitor how the energy influences the overall orbital structure of our dynamical system using a continuous spectrum of Jacobi constants rather than few discrete values. In Fig. \ref{xC}(a-d) we present the orbital structure of the $(x,C)$ plane for four values of the radiation pressure factor when $C \in [-6,6]$, while in Fig. \ref{xCt}(a-d) the distribution of the corresponding escape and collision times of the orbits is depicted. The black solid line in Fig. \ref{xC}(a-d) is the limiting curve which distinguishes between regions of allowed and forbidden motion and is defined as
\begin{equation}
f_L(x,C) = 2\Omega(x,y = 0) = C.
\label{zvc}
\end{equation}

We can observe the presence of several types of regular orbits around the two primary bodies. Being more precise, on both sides of the primaries we identify stability islands corresponding to both direct (counterclockwise) and retrograde (clockwise) quasi-periodic orbits. It is seen that a large portion of the exterior region, that is for $x < x(L_3)$ and $x > x(L_2)$, a large portion of the $(x,C)$ plane is covered by initial conditions of escaping orbits however, at the left-hand side of the same plane two stability islands of regular orbits that circulate around both primaries are observed. Additional numerical calculations reveal that for much lower values of $x$ $(x < 5)$ these two stability islands are joined and form a crescent-like shape. We also see that collisional basins to radiating primary 1 leak outside the interior region, mainly outside $L_3$, and create complicated spiral shapes in the exterior region. On the other hand, the thin bands represent initial conditions of orbits that collide with primary body 2 are much more confined. It should be pointed out that in the blow-ups of the diagrams several additional very small islands of stability have been identified\footnote{An infinite number of regions of (stable) quasi-periodic (or small scale chaotic) motion is expected from classical chaos theory.}.

As the value of the radiation pressure factor decreases (which means that the radiation of the primary body increases) the structure of the $(x,C)$ planes exhibits the following changes: (i) The collisional basins to radiating primary 1 gradually weakens; (ii) The area of the stability islands around primary 1 reduces. According to Broucke's classification \citep{B68} the periodic orbits around the primaries belong to the families $C$ (at the left side of the primary) and $H_1$ (at the right side of the primary), while \citet{SS00} proved for the planar Hill's problem that the stability regions of the $C$ family are  more stable than those of the $H_1$ family; (iii) The stability islands around primary body 2 are almost unperturbed, at least in the interval $C \in [-6, -6]$, by the decrease on the value of the radiation pressure factor. The phenomenon that stability islands can appear and disappear as a dynamical parameter is changed has also been reported in earlier paper (e.g., \citet{BBS06,dAT14}); (iv) The collisional basins to primary 2 increases as the value of the radiation pressure factor decreases; (vi) The two stability islands in the exterior region come closer as the intensity of the radiation of primary 1 increases; (vii) Another interesting phenomenon is the fact that as the intensity of the radiation of primary body 1 increases the fractality of the $(x,C)$ plane reduces and the boundaries between escaping and collisional basins appear to become smoother. It should be emphasized that the fractality of the structures was not measured by computing the corresponding fractal dimension. When we state that an area is fractal we mean that it has a fractal-like geometry.

It would be very informative to monitor the evolution of the percentages of the different types of orbits as a function of the Jacobi constant $C$ for the $(x,C)$ planes shown in Figs. \ref{xC}(a-d). Our results are presented in Figs. \ref{pxC}(a-d). We see that in all four cases the percentages display similar patterns, so we are going to explain only the first case shown in Fig. \ref{pxC}a where $q = 0.9$. For $C < -5$ escaping orbits cover all the available space however their rate gradually reduces until about 45\% for $C = -2.5$. For $C > -2.5$ it suddenly increases and in the interval $[-1.5, 3]$ escaping orbits dominate with rates above 80\%. The evolution of the percentage of regular bounded orbits displays an exact opposite evolution with respect to the pattern of escaping orbits. In particular, in the interval $[-1.5, 3]$ the percentage of regular orbits fluctuate at relatively low values below 12\%, while for $C < -2$ and $C > 3$ on the other hand, two peaks are observed at about 55\% and 75\%, respectively. The percentages of collisional orbits to primaries 1 and 2 have a monotone behaviour with almost zero values for $C < -2$ and $C > 3.5$, while in the interval $[-2, 3.5]$ their rates fluctuate at extremely low values below 5\%. Thus we may argue that the most interesting interval of values of the Jacobi constant $C$ is the interval $[-2, 3.5]$. Inspecting all four sub-panels it becomes evident that this particular interval becomes smaller and smaller as the value of the radiation pressure factor decreases and the intensity of the radiation of the more massive primary becomes stronger. Therefore one may reasonably conclude that the radiation pressure factor practically does not drastically affect the actual percentages of the different types of orbits. All it does is to shift the energy intervals at which the different patterns appear.

Another interesting aspect would be to reveal how the radiation pressure factor $q$ influences the escape as well as the collision time of orbits. The evolution of the average value of the escape time $< t_{\rm esc} >$ of orbits as a function of the value of the Jacobi constant $C$ is given in Fig. \ref{tesc}. It is evident, especially in the energy interval $[-1.5, 1.5]$, that as the value of the radiation pressure decreases the escape rates of orbits are reduced. In the same vein in Fig. \ref{tcol}a we present the evolution of the average collision time $< t_{\rm col} >$ of orbits that collide to the radiating primary 1. Once more we observe that the collision time of orbits in general terms decreases with decreasing value of $q$. Additional numerical calculations, shown in Fig. \ref{tcol}b, indicate that the radiation pressure factor also influences the collision rates of orbits which collide to non-radiating primary 2. In particular, in the range $[-1, 2.5]$ the collision rates to primary 2 are reduced with decreasing $q$.

Before closing this section we would like to add that the particular value of the mass ratio $\mu$ does not really change the qualitative nature of the numerical outcomes presented in this section. Indeed after conducting some additional calculations with larger and lower values of $\mu$ we concluded to the same results. The parameters which mostly influence the orbital dynamics are the total orbital energy and of course the radiation pressure factor.

\section{Discussion and conclusions}
\label{conc}

The main scope of this numerical investigation was to unveil how the radiation pressure factor influences the character of orbits in the classical planar circular photogravitational restricted three-body problem. After conducting an extensive and thorough numerical investigation we managed to distinguish between bounded, escaping and collisional orbits and we also located the basins of escape and collision, finding also correlations with the corresponding escape and collision times. Our numerical results strongly suggest that the radiation pressure factor plays a very important role in the nature of the test's body motion under the gravitational field of the two primaries. To our knowledge, this is the first detailed and systematic numerical analysis on the influence of the radiation pressure factor on the character of orbits and this is exactly the novelty and the contribution of the current work.

For several values of the radiation pressure factor in the last four Hill's regions configurations we defined dense uniform grids of $1024 \times 1024$ initial conditions regularly distributed on the $\dot{\phi} < 0$ part of the configuration $(x,y)$ plane inside the area allowed by the value of the Jacobi constant (or in other words by the value of the total orbital energy). All orbits were launched with initial conditions inside the scattering region, which in our case was a square grid with $-2\leq x,y \leq 2$. For the numerical integration of the orbits in each type of grid, we needed about between 12 hours and 6 days of CPU time on a Pentium Dual-Core 2.2 GHz PC, depending on the escape and collisional rates of orbits in each case. For each initial condition, the maximum time of the numerical integration was set to be equal to $10^4$ time units however, when a particle escaped or collided with one of the two primaries the numerical integration was effectively ended and proceeded to the next available initial condition.

In this study we provide quantitative information regarding the escape and collisional dynamics in the photogravitational restricted three-body problem. The main outcomes of our numerical research can be summarized as follows:
\begin{enumerate}
 \item We found that for $q < q^{*}$ when the neck around $L_2$ is open, the throat around $L_1$ is still closed. For very low values of the radiation pressure factor the allowed area of motion around the center of the more massive primary is very small, while when the throat around $L_1$ opens the banana-shaped forbidden regions surround only the radiating primary.
 \item It was observed that as the value of the radiation pressure factor decreases the area of the stability islands around primary 1 is reduced, while the area of the stability islands around non-radiating primary 2 remains almost unperturbed.
 \item As the radiation pressure of primary 1 increases the initial conditions in the exterior region of collisional orbits to primary 1 appear only at relatively low values of the Jacobi constant, or in other words at high values of the total orbital energy. At the same time the collisional basins to primary 2 in the exterior region weaken with reducing $C$.
 \item It was detected that the collisional basins to primary 1 are reduced with decreasing value of the radiation pressure factor, while the collisional basins to primary 2 are strengthened as the intensity of the radiation pressure increases.
 \item We presented numerical evidence that the radiation pressure factor also influences the escape as well as the collision average time of the orbits. In particular, both types of average times are reduced as the value of the value of the radiation pressure factor decreases.
 \item Our calculations reveal that as the primary body 1 becomes more and more radiating the fractality of the planes is reduced and the boundaries between the bounded, escape and collisional basins appear to become smoother.
\end{enumerate}

Judging by the detailed and novel outcomes we may say that our task has been successfully completed. We hope that the present numerical analysis and the corresponding results to be useful in the field of escape dynamics in the photogravitational restricted three-body problem. The results as well as the conclusions of the present research are considered, as an initial effort and also as a promising step in the task of understanding the escape mechanism of orbits in this interesting version of the classical three-body problem. Taking into account that our outcomes are encouraging, it is in our future plans to properly modify our dynamical model in order to expand our investigation into three dimensions and explore the entire six-dimensional phase thus revealing the influence of the radiation pressure factor on the orbital structure.

\section*{Acknowledgments}

I would like to express my warmest thanks to the anonymous referee for the careful reading of the manuscript and for all the apt suggestions and comments which allowed us to improve both the quality and the clarity of the paper.

\end{document}